\documentclass[11pt]{article}
\usepackage[dvipdfmx]{color}
\usepackage{graphicx}
\usepackage{xcolor}
  \usepackage{amsthm,amsfonts}
  \usepackage{amsmath}
\usepackage{slashed}
\usepackage{ulem}
\usepackage{cite}

\newcommand{\bea}   {\begin{eqnarray}}
\newcommand{\eea}   {\end{eqnarray}}

\begin{document}
\renewcommand{\thefootnote}{\fnsymbol{footnote}}

\thispagestyle{empty}

\title{${\mathbb Z}_2\times {\mathbb Z}_2$-graded mechanics: the classical theory}
\author{N. Aizawa\thanks{{E-mail: {\it aizawa@p.s.osakafu-u.ac.jp}}}, \quad Z. Kuznetsova\thanks{{E-mail: {\it zhanna.kuznetsova@ufabc.edu.br}}}\quad and\quad
F.
Toppan\thanks{{E-mail: {\it toppan@cbpf.br}}}
\\
\\
}
\maketitle

\centerline{$^{\ast}$ {\it Department of Physical Sciences, Graduate School of Science,}}{\centerline{\it  Osaka Prefecture University, Nakamozu Campus,}}{\centerline{\it Sakai, Osaka 599-8531 Japan.}}
\centerline{$^{\dag}$ {\it UFABC, Av. dos Estados 5001, Bangu,}}\centerline{\it { cep 09210-580, Santo Andr\'e (SP), Brazil.}}
{\centerline{$^{\ddag}$ 
{\it CBPF, Rua Dr. Xavier Sigaud 150, Urca,}}\centerline{\it{
cep 22290-180, Rio de Janeiro (RJ), Brazil.}}
~\\
\maketitle
\begin{abstract}
${\mathbb Z}_2\times {\mathbb Z}_2$-graded mechanics admits four types of particles: ordinary bosons, two classes of fermions (fermions belonging to different classes commute among each other) and exotic bosons. In this paper we 
construct the basic ${\mathbb Z}_2\times {\mathbb Z}_2$-graded worldline multiplets (extending the cases of one-dimensional supersymmetry) and compute, based on a general scheme, their invariant classical actions and worldline sigma-models. The four basic multiplets contain two bosons and two fermions. They are $(2,2,0)$, with two propagating bosons and two propagating fermions, $(1,2,1)_{[00]}$ (the ordinary boson is  propagating, while the exotic boson is an auxiliary field), $(1,2,1)_{[11]}$
(the converse case, the exotic boson is propagating, while the ordinary boson is an auxiliary field) and, finally, $(0,2,2)$ with two bosonic auxiliary fields. Classical actions invariant under the ${\mathbb Z}_2\times {\mathbb Z}_2$-graded superalgebra are constructed for both single multiplets and interacting multiplets.  Furthermore, scale-invariant actions can possess a full ${\mathbb Z}_2\times {\mathbb Z}_2$-graded conformal invariance spanned by
$10$ generators and containing an $sl(2)$ subalgebra.
\end{abstract}
\vfill
\rightline{CBPF-NF-002/20}
\newpage

\section{Introduction}

In this paper we present the framework to construct ${\mathbb Z}_2\times {\mathbb Z}_2$-graded mechanics as a one-dimensional classical theory. We extend to the ${\mathbb Z}_2\times {\mathbb Z}_2$-graded setting the results and approach employed for ordinary worldline supersymmetric \textcolor{black}{\cite{PaTo,KRT}} and superconformal \textcolor{black}{\cite{KT}} mechanics at the classical level.
A separate paper is devoted to the quantization of the models here introduced.\par
The theory under  consideration possesses four types of time-dependent fields: ordinary bosons, two classes of fermions (fermions belonging to different classes commute among themselves) and exotic bosons which anticommute with the fermions of both classes.\par
Before discussing ${\mathbb Z}_2\times {\mathbb Z}_2$-graded theories, we briefly sketch the list of the main results of this work. We introduce at first the basic $4$ component-field multiplets and their respective $D$-module representations
(realized by $4\times 4$ matrix differential operators) of both ${\mathbb Z}_2\times {\mathbb Z}_2$-graded superalgebra and superconformal  algebra.  These results should be compared with the supermultiplets of the ${\cal N}=2$ worldline supersymmetry. In that case  the basic multiplets are \textcolor{black}{\cite{PaTo}} the ``chiral" supermultiplet \cite{{GatRan},{GatRan2}}, also known in the literature \cite{BBMO} as the ``root" supermultiplet and denoted as ``$(2,2,0)$" (in physical applications it produces $2$ propagating bosons, $2$ propagating fermions and no auxiliary field), and the real supermultiplet $(1,2,1)$ with one bosonic auxiliary field. The notation ``$(0,2,2)$" is sometimes used to denote the supermultiplet with two bosonic auxiliary fields.  In  application to the
${\mathbb Z}_2\times {\mathbb Z}_2$-graded superalgebra, similar notations can be applied, but one has to discriminate between the two subcases $(1,2,1)_{[00]}$ and $(1,2,1)_{[11]}$, where the suffix denotes which bosonic field is propagating, either the ordinary one (the $(1,2,1)_{[00]}$ multiplet), or the exotic one (the $(1,2,1)_{[11]}$ multiplet). In the following we construct classical invariant actions in the Lagrangian framework for both single basic multiplets and several interacting basic multiplets. The construction relies on the ${\mathbb Z}_2\times {\mathbb Z}_2$-graded Leibniz property satisfied by the matrix differential operators closing the ${\mathbb Z}_2\times {\mathbb Z}_2$-graded superalgebra. {\textcolor{black}{Following the approach of \cite{KT}}}, ${\mathbb Z}_2\times {\mathbb Z}_2$-graded superconformal invariant actions are obtained by requiring invariance under $K$, the conformal counterpart of the time-translation generator $H$ {\textcolor{black}{(for a review of ordinary superconformal mechanics see \cite{FIL}).}} \par

\textcolor{black}{
The present work is motivated by the recently introduced ${\mathbb Z}_2\times {\mathbb Z}_2$-graded} {\textcolor{black}{analogue}} \textcolor{black}{of supersymmetric and superconformal quantum mechanics \cite{Bru,BruDup,NaAmaDoi,NaAmaDoi2}. 
}
\textcolor{black}{
It consists of models of one particle quantum mechanics on a real line whose symmetries are described by ${\mathbb Z}_2\times{\mathbb Z}_2$-graded Lie superalgebras. 
Peculiar to these models is the fact that their supercharges close with commutators, instead of anticommutators; this feature, which reflects the ${\mathbb Z}_2\times{\mathbb Z}_2$-grading, impies that two types of fermions commute with each other. One should also add that in ${\mathbb Z}_2\times{\mathbb Z}_2$-graded models central elements appear naturally.
Due to properties of this type, the physical meaning of the $ {\mathbb Z}_2\times{\mathbb Z}_2$-graded models has yet to be properly understood and clarified.}
{\par
}\textcolor{black}{
We recall that supersymmetry algebra with central extension naturally appears in higher dimensional Dirac actions with curved extra dimension, see \cite{Ueba}, while commuting fermions also appear in the dual double field theory \cite{BHPR} (see also \cite{CKRS} and \cite{BruIbar} for the relation to higher grading geometry). 
Therefore, one could expect that the $ {\mathbb Z}_2\times{\mathbb Z}_2$-graded quantum systems possess some physical relevance, at least in nonrelativistic or anyonic physics, where the spin-statistics connection does not necessarily hold.  Obviously, a thorough understanding of these quantum systems is highly desirable. 
}

\textcolor{black}{
For a better understanding of the results in \cite{Bru,BruDup,NaAmaDoi,NaAmaDoi2}, one may pose the following question: 
is it possible to recover and extend  these given models  by quantizing some classical system? 
This question relies on a more essential one, namely, are there examples of classical systems invariant under the $ {\mathbb Z}_2\times{\mathbb Z}_2$-graded supersymmetric transformations? And, if this is indeed so, which
are the $ {\mathbb Z}_2\times{\mathbb Z}_2$-graded supersymmetric transformations? 
In the present work we positively answer the last two questions about classical systems. The answer to the first question about quantization is postponed to a separate paper, see \cite{akt}. 
}

\textcolor{black}{
After the introduction  \textcolor{black}{\cite{Ree,rw1,rw2,sch}} of ${\mathbb Z}_2\times{\mathbb Z}_2$-graded Lie superalgebras, a long history of investigations of $ {\mathbb Z}_2\times{\mathbb Z}_2$ (and higher) graded symmetries in physical problems, which recently attracted a renewed interest, began. Several works considered enlarged symmetries in various contexts such as
extensions of spacetime symmetries (beyond ordinary de-Sitter and Poincar\'e algebras),  supergravity theory,
quasi-spin formalism, parastatistics and non-commutative geometry, see \cite{lr,vas,jyw,zhe,Toro1,Toro2,tol2,tol,BruDup2,Me}. 
It was also recently revealed that the symmetries of the L\'evy-Leblond equation, which is a nonrelativistic quantum mechanical wave equation for spin 1/2 particles, are given by a ${\mathbb Z}_2\times {\mathbb Z}_2$-graded superalgebra \cite{aktt1,aktt2}.	
}

\textcolor{black}{
Various mathematical studies of algebraic and geometric aspects of higher graded superalgebras have also been undertaken since their introduction. In this respect one of the hot topics is the geometry of higher graded manifolds, which is  an extension of the geometry of supermanifolds. 
For those mathematical works the reader may consult the references in \cite{AiIsSe}.
}

The scheme of the paper is as follows: in Section {\bf 2} we review some basic features of ${\mathbb Z}_2\times{\mathbb Z}_2$-graded superalgebras and their $4\times 4$ real matrices representations. In Section {\bf 3} we introduce the $D$-module representations of the ${\mathbb Z}_2\times {\mathbb Z}_2$-graded superalgebra. In Section {\bf 4} we present  the superconformal extension of the $D$-module representations. The construction of ${\mathbb Z}_2\times {\mathbb Z}_2$-graded classical invariant actions is given in Section {\bf 5}.  {\textcolor{black}{We introduce in Appendix {\bf A} the scaling dimensions of the ${\mathbb Z}_2\times{\mathbb Z}_2$-graded generators and  of the component fields entering the multiplets. We present in Appendix {\bf B} a derivation (that can be extended to the corresponding ${\mathbb Z}_2\times {\mathbb Z}_2$-graded case) of the ${\cal N}=2$ supersymmetric action for the real superfield. }} In the Conclusions we comment about future developments and the quantization of the models.

\section{On ${\mathbb Z}_2\times{\mathbb Z}_2$-graded matrices}
 
We recall at first the definition \textcolor{black}{\cite{Ree,rw1,rw2,sch}} of a ${\mathbb Z}_2\times{\mathbb Z}_2$-graded Lie superalgebra ${\cal G}$.\par  
It admits the decomposition
\bea
{\cal G}&=& {\cal G}_{00}\oplus {\cal G}_{10}\oplus{\cal G}_{01}\oplus {\cal G}_{11}
\eea
and is endowed with an operation $[\cdot,\cdot\}: {\cal G}\times {\cal G}\rightarrow {\cal G}$ which satisfies the following properties
for any $g_a\in {\cal G}_{{\vec{\alpha}}}$, 
\par
{1)} the ${\mathbb Z}_2\times{\mathbb Z}_2$-graded (anti)commutation relations
\bea\label{anticommrel}
[g_a,g_b\} &=& g_ag_b-(-1)^{{\vec{\alpha}}\cdot{\vec{\beta}}}g_bg_a,
\eea
\par 
2) the ${\mathbb Z}_2\times{\mathbb Z}_2$-graded Jacobi identity
\bea
(-1)^{{\vec{\gamma}}\cdot{\vec{\alpha}}}[g_a,[g_b,g_c\} \}+
(-1)^{{\vec{\alpha}}\cdot{\vec{\beta}}}[g_b,[g_c,g_a\} \}+
(-1)^{{\vec{\beta}}\cdot{\vec{\gamma}}}[g_c,[g_a,g_b\} \}&=& 0.
\eea
In the above formulas $g_a, g_b, g_c$ respectively belong to the sectors ${\cal G}_{{\vec{\alpha}}},{\cal G}_{{\vec{\beta}}},{\cal G}_{{\vec{\gamma}}}$, where ${\vec{\alpha}}=(\alpha_1,\alpha_2)$ for $\alpha_{1,2}=0,1$
and ${\cal G}_{\vec\alpha}\equiv {\cal G}_{\alpha_1\alpha_2}$
(similar expressions hold for ${\vec{\beta}}$ and ${\vec{\gamma}}$). \par The scalar product ${\vec{\alpha}}\cdot{\vec{\beta}}$ is defined as
\bea\label{scalarproduct}
{\vec{\alpha}}\cdot{\vec{\beta}}&=& \alpha_1\beta_1+\alpha_2\beta_2.
\eea
Finally,
$ [g_a,g_b\}\in {\cal G}_{{\vec{\alpha}}+{\vec{\beta}}}$ where the vector sum is defined ${\textrm{mod}}~ 2$.\par
 A ${\mathbb Z}_2\times{\mathbb Z}_2$-graded Lie superalgebra ${\cal G}$  can be realized by $4\times 4$, real matrices which can be accommodated into the ${\cal G}_{ij}$ sectors of ${\cal G}$ according to
\bea\label{gradedmatrices}
 {\cal G}_{00}= \left(\begin{array}{cccc}\ast&0&0&0\\0&\ast&0&0\\0&0&\ast&0\\0&0&0&\ast\end{array}\right), &\quad&
 {\cal G}_{11}= \left(\begin{array}{cccc}0&\ast&0&0\\\ast&0&0&0\\0&0&0&\ast\\0&0&\ast&0\end{array}\right),
\nonumber\\
{\cal G}_{10}= \left(\begin{array}{cccc}0&0&\ast&0\\0&0&0&\ast\\\ast&0&0&0\\0&\ast&0&0\end{array}\right),
 &\quad&
 {\cal G}_{01}= \left(\begin{array}{cccc}0&0&0&\ast\\ 0&0&\ast&0\\ 0&\ast&0&0\\ \ast&0&0&0\end{array}\right),
\eea
 where the ``$\ast$" symbol denotes the non-vanishing real entries.\par
The matrix generators spanning each ${\mathbb Z}_2\times{\mathbb Z}_2$-graded sector can be expressed as
tensor products of the $4$ real, $2\times 2$ split-quaternion matrices $I,X,Y,A$ (see \textcolor{black}{\cite{aktt1}}) given by  
{\footnotesize{
\bea\label{splitquat}
&I= \left(\begin{array}{cc}1&0\\0&1\end{array}\right),\quad X= \left(\begin{array}{cc}1&0\\0&-1\end{array}\right),\quad
Y= \left(\begin{array}{cc}0&1\\1&0\end{array}\right),\quad
A= \left(\begin{array}{cc}0&1\\-1&0\end{array}\right).&
\eea
}}
We have
\bea
{\cal G}_{00}&:& \quad I\otimes I, \quad~ I\otimes X,\quad X\otimes I,\quad X\otimes X,\nonumber\\
{\cal G}_{11}&:& \quad I\otimes Y, \quad ~ I\otimes A,\quad X\otimes Y,\quad X\otimes A,\nonumber\\
{\cal G}_{10}&:& \quad Y\otimes I, \quad Y\otimes X,\quad A\otimes I,\quad A\otimes X,\nonumber\\
{\cal G}_{01}&:& \quad Y\otimes Y, \quad Y\otimes A,\quad A\otimes Y,\quad A\otimes A.
\eea
Up to an overall normalization, the most general Hermitian matrices with real coefficients and respectively belonging to the  ${\cal G}_{10}$ and ${\cal G}_{01}$ sectors
are
\bea
Q_{10}= \cos \alpha~ Y\otimes I+\sin\alpha ~Y\otimes X, &\quad&
Q_{01} = \cos \beta ~Y\otimes Y +\sin\beta ~A\otimes A,
\eea
where $\alpha,\beta$ are arbitrary angles. If $\alpha,\beta\neq n\frac{\pi}{2}$ for $n\in {\mathbb Z}$, then both $Q_{10}^2\neq {\mathbb I}_4$ and $Q_{01}^2\neq {\mathbb I}_4$ (${\mathbb I}_4=I\otimes I$ is the $4\times 4$ identity matrix). \par
Working under the assumption that $Q_{10}^2=Q_{01}^2={\mathbb I}_4$, the following choices for $Q_{10}$ are admissible.  Either $Q_{10}=\pm Y\otimes I$ or $Q_{10}=\pm Y\otimes X$. Up to the overall sign and without loss of generality (the second choice being recovered from the first one via a similarity transformation) we can set
\bea
Q_{10}&=& Y\otimes I.
\eea
This position implies, up to a sign, two possible choices for $Q_{01}$: 
\bea 
{\textrm {either}}\quad Q_{01}^A =Y\otimes Y & {\textrm{or}}&
Q_{01}^B = A\otimes A.
\eea
We explore the consequences of each one of these choices.\par
~\par
By assuming choice  $A$ one can introduce\par 
~\par
$1_A$) a ${\mathbb Z}_2$-graded superalgebra  spanned by the two odd generators $Q_{10}, ~Q_{01}^A$ and the two even generators ${\mathbb I}_4, ~ W= I\otimes Y$, which is defined by the non-vanishing (anti)commutators
\bea
\{Q_{10},Q_{10}\}=\{Q_{01}^A,Q_{01}^A\}= 2\cdot{\mathbb I}_4, &&\{Q_{10},Q_{01}^A\}= 2W.
\eea
Since $\{Q_{01},Q_{10}^A\}=W\neq 0$, this superalgebra does not correspond to the ordinary ${\cal N}=2$ worldline supersymmetry;\par
~\par
$2_A$) a ${\mathbb Z}_2\times{\mathbb Z}_2$-graded superalgebra defined by the (anti)commutators
\bea
\relax &\{Q_{10},Q_{10}\}=\{Q_{01}^A,Q_{01}^A\}= 2\cdot{\mathbb I}_4, \quad [Q_{10},Q_{01}^A]=0.&
\eea
A non-vanishing ${\cal G}_{11}$  sector  can be introduced by adding an operator $Z\in {\cal G}_{11}$, given by
\bea
Z &=& \epsilon I\otimes Y + r X\otimes Y, \quad {\textrm{with}} \quad \epsilon=0,1, \quad r \in {\mathbb R}.
\eea
It follows that
\bea
\{Q_{01}^A, Z\} = 2\epsilon Q_{10}, &\quad& \{Q_{10}, Z\} = 2\epsilon Q_{01}^A.
\eea
\par
~\par
By assuming choice $B$ one can introduce\par
~\par
$1_B$) a ${\mathbb Z}_2$-graded superalgebra given by the (anti)commutators
\bea
&\{Q_{10},Q_{10}\}=\{Q_{01}^B,Q_{01}^B\}= 2\cdot{\mathbb I}_4, \quad \{Q_{10},Q_{01}^B\}= 0.&
\eea
This superalgebra corresponds to the ${\cal N}=2$ worldline supersymmetry algebra;\par
~\par
$2_B$) a ${\mathbb Z}_2\times{\mathbb Z}_2$-graded superalgebra spanned by the generators 
${\mathbb I}_4,~Q_{10}, ~Q_{01}^B, ~Z=X\otimes A$ and defined by the non-vanishing (anti)commutators
\bea\label{z2z2const}
\relax \{Q_{10},Q_{10}\}=\{Q_{01}^B,Q_{01}^B\}= 2\cdot{\mathbb I}_4,&& [Q_{10}, Q_{01}^B] = -2Z.
\eea

\section{$D$-module representations and supermultiplets of the ${\mathbb Z}_2\times{\mathbb Z}_2$-graded worldline superalgebra}

In analogy with the $D$-module representations \textcolor{black}{\cite{PaTo, KRT}} of ordinary worldline supersymmetry, the $D$-module representations of the ${\mathbb Z}_2\times{\mathbb Z}_2$-graded superalgebra are obtained by replacing the constant matrices with real coefficients (as those entering formula (\ref{z2z2const})) with differential matrix operators. Since these matrices with differential entries are
$4\times 4$, a total number of four, time-dependent, real fields are required. In the following the time coordinate is denoted by ``$\tau$". In the construction of the  ${\mathbb Z}_2\times{\mathbb Z}_2$-graded worldline multiplets it is convenient to assume the reality conditions on all four fields. On the other hand, 
in the construction of sigma-models, it is sometimes convenient to assume the fields to be Hermitian and work with the Wick-rotated time coordinate $t$, given by
$t=i\tau$, so that  $\partial_t=-i\partial_\tau$. One can easily go back and forth from the ``Euclidean time" $\tau$ to the
real time $t$ through Wick rotation. Some fields which are real in the Euclidean time version become imaginary in the real time formalism.\par
The reality condition
is associated with the complex conjugation (denoted by ``$\ast$"). The hermiticity condition is associated with the adjoint operator
(``$\dagger$") given by a complex conjugation and a transposition (denoted by ``$T$"). 
The operations satisfy 
\bea
&(A^\ast)^\ast=(A^T)^T= (A^\dagger)^\dagger=A, \quad (AB)^\ast= A^\ast B^\ast,\quad (AB)^T=B^TA^T,\quad (AB)^\dagger = B^\dagger A^\dagger.&
\eea
They are interrelated through
\bea
A^\dagger = (A^\ast)^T.
\eea
Without loss of generality, all formulas in the paper are presented for the Euclidean time $\tau$.\par
A given multiplet $m^T= (x(\tau), z(\tau),\psi(\tau), \xi(\tau))$ of time-dependent fields belongs to a ${\mathbb Z}_2\times{\mathbb Z}_2$-graded vector space ${\cal V}$,
\bea
&x(\tau)\in{\cal V}_{00}, \quad z(\tau)\in {\cal V}_{11}, \quad\psi(\tau)\in{\cal V}_{10},\quad\xi(\tau)\in{\cal V}_{01},&
\eea
such that its grading is consistent with
the ${\mathbb Z}_2\times{\mathbb Z}_2$-grading of the differential operators.\par
A field belonging to ${\cal V}_{\epsilon_1\epsilon_2}$  is called ``even" (respectively ``odd") if the sum $\epsilon_1+\epsilon_2$ ($\epsilon_1+\epsilon_2=0~{\textrm{mod}} ~2$ or $\epsilon_1+\epsilon_2=1~{\textrm{mod}} ~2$) is even (odd).\par
Four types of multiplets are encountered:
\bea\label{fourmultiplets}
&(2,2,0), \qquad (1,2,1)_{[00]},\qquad (1,2,1)_{[11]},\qquad (0,2,2).&
\eea
The first multiplet corresponds to the ``root" multiplet with two propagating even fields and two propagating odd fields. The $(1,2,1)$ multiplets, just like the corresponding ${\cal N}=2$ worldline supermultiplet, correspond to one even propagating field,
two odd propagating fields and one even auxiliary field. An extra piece of information has to be added. The suffix
$[00]$ specifies that the even propagating field is the ordinary boson, while the $[11]$ suffix specifies that the even 
propagating field is the exotic boson. Finally, the $(0,2,2)$ multiplet corresponds to the case of two propagating odd and two auxiliary even fields. \par As for the worldline supersymmetry, the multiplets  $(1,2,1)_{[00]},~(1,2,1)_{[11]},~ (0,2,2)$ are obtained from the root multiplet $(2,2,0)$ via a dressing transformation.\par
The $D$-module representations associated with the ${\mathbb Z}_2\times{\mathbb Z}_2$-graded Lie superalgebra (\ref{z2z2const}) are presented in the two following subsections. The operators are $H\in {\cal G}_{00}, ~Z\in{\cal G}_{11}, ~Q_{10}\in {\cal G}_{10}, ~Q_{01}\in {\cal G}_{01}$. The operator $H$ is the generator of the time translation. It commutes with all algebra generators and replaces  the identity ${\mathbb I}_4$ in (\ref{z2z2const}). The unnecessary label ``$B$" is dropped in the definition of the $Q_{01}$ operator. \par The
${\mathbb Z}_2\times{\mathbb Z}_2$-graded superalgebra is defined by the non-vanishing (anti)commutators
\bea\label{z2z2super}
\{Q_{10},Q_{10}\}=\{Q_{01},Q_{01}\}=2H,  &\quad& [Q_{10},Q_{01}] = -2Z.
\eea

\subsection{The root multiplet}

The differential operators associated with the $(2,2,0)$ root multiplet are
{
\footnotesize{
\bea\label{root}
H= \left(\begin{array}{cccc}\partial_{\tau}&0&0&0\\0&\partial_{\tau}&0&0\\0&0&\partial_{\tau}&0\\ 
0&0&0&\partial_{\tau}\end{array}\right), &\quad&
Z= \left(\begin{array}{cccc}0&\partial_{\tau}&0&0\\-\partial_{\tau}&0&0&0\\0&0&0&-\partial_{\tau}
\\ 0&0&\partial_{\tau}&0\end{array}\right),\nonumber\\
Q_{10}= \left(\begin{array}{cccc}0&0&1&0\\0&0&0&1\\ \partial_{\tau}&0&0&0\\ 
0&\partial_{\tau}&0&0\end{array}\right), &\quad&
Q_{01}= \left(\begin{array}{cccc}0&0&0&1\\0&0&-1&0\\0&-\partial_{\tau}&0&0
\\ \partial_{\tau}&0&0&0\end{array}\right).
\eea
}}
The corresponding transformations of the component fields are (for simplicity here and in the following it is not 
needed to report the action of $H$,  being just a time derivative)
\bea\label{rootfieldtransf}
&\begin{array}{cccc}Q_{10}x={\psi},~~&Q_{10}z={\xi},~~&Q_{10}\psi={\dot x},~~&Q_{10}\xi={\dot z},~~\\
Q_{01}x={\xi},~~&Q_{01}z=-{\psi},~~&Q_{01}\psi=-{\dot z},~~&Q_{01}\xi={\dot x},~~\\
~~Z x={\dot z},~~&~~Zz=-{\dot x},~~&~~Z \psi=-{\dot\xi},~~&~~Z \xi={\dot\psi}.~~
\end{array}&
\eea

\subsection{The dressed multiplets}

Following the derivation \textcolor{black}{\cite{PaTo, KRT}} of the worldline supermultiplets, the remaining ${\mathbb Z}_2\times {\mathbb Z}_2$-graded multiplets are obtained from the operators given in (\ref{root}) and associated with the root multiplet, by applying the dressing transformation
\bea
{\frak M}&\mapsto{\frak  M}'={\frak D} {\frak M} {\frak D}^{-1}.
\eea
In the above formula  ${\frak M}$ denotes any operator in (\ref{root}), while ${\frak D}$ is a differential diagonal operator. The three consistent choices for ${\frak D}$,
\bea\label{dressingmatrices}
&{\frak D}_1 =diag(\partial_\tau, 1,1,1), \quad{\frak D}_2=diag(1,\partial_\tau,1,1),\quad {\frak D}_3=diag(\partial_\tau,\partial_\tau,1,1),
\eea
are such that the transformed operators ${\frak M}'$, despite the presence of ${\frak D}^{-1}$ in the right hand side, remain differential operators. They correspond to the $D$-module representations respectively acting on the $(1,2,1)_{[11]}$, $ (1,2,1)_{[00]}$ and $(0,2,2)$ multiplets. They are given by:
\par
~
\par
{\it i}) for the $(1,2,1)_{[11]}$ multiplet the $D$-module representation is
{
\footnotesize{
\bea\label{12111rep}
H= \left(\begin{array}{cccc}\partial_{\tau}&0&0&0\\0&\partial_{\tau}&0&0\\0&0&\partial_{\tau}&0\\ 
0&0&0&\partial_{\tau}\end{array}\right), &\quad&
Z= \left(\begin{array}{cccc}0&\partial_{\tau}^2&0&0\\-1&0&0&0\\0&0&0&-\partial_{\tau}
\\ 0&0&\partial_{\tau}&0\end{array}\right),\nonumber\\
Q_{10}= \left(\begin{array}{cccc}0&0&\partial_\tau&0\\0&0&0&1\\ 1&0&0&0\\ 
0&\partial_{\tau}&0&0\end{array}\right), &\quad&
Q_{01}= \left(\begin{array}{cccc}0&0&0&\partial_{\tau}\\0&0&-1&0\\0&-\partial_{\tau}&0&0
\\ 1&0&0&0\end{array}\right).
\eea
}
}
The corresponding transformations of the component fields are
\bea\label{12111transf}
&\begin{array}{cccc}Q_{10}x={\dot{\psi}},~~&Q_{10}z={\xi},~~&Q_{10}\psi={x},~~&Q_{10}\xi={\dot z},~~\\
Q_{01}x={\dot{\xi}},~~&Q_{01}z=-{\psi},~~&Q_{01}\psi=-{\dot z},~~&Q_{01}\xi={x},~~\\
~~Z x={\ddot z},~~&~~Zz=-{x},~~&~~Z \psi=-{\dot\xi},~~&~~Z \xi={\dot\psi};~~
\end{array}&
\eea
\par
~\par
{\it ii}) for the $(1,2,1)_{[00]}$ multiplet the $D$-module representation is
{
\footnotesize{
\bea\label{12100rep}
H= \left(\begin{array}{cccc}\partial_{\tau}&0&0&0\\0&\partial_{\tau}&0&0\\0&0&\partial_{\tau}&0\\ 
0&0&0&\partial_{\tau}\end{array}\right), &\quad&
Z= \left(\begin{array}{cccc}0&1&0&0\\-\partial_{\tau}^2&0&0&0\\0&0&0&-\partial_{\tau}
\\ 0&0&\partial_{\tau}&0\end{array}\right),\nonumber\\
Q_{10}= \left(\begin{array}{cccc}0&0&1&0\\0&0&0&\partial_{\tau}\\ \partial_{\tau}&0&0&0\\ 
0&1&0&0\end{array}\right), &\quad&
Q_{01}= \left(\begin{array}{cccc}0&0&0&1\\0&0&-\partial_\tau&0\\0&-1&0&0
\\ \partial_{\tau}&0&0&0\end{array}\right).
\eea
}
}
The corresponding transformations of the component fields are
\bea\label{12100transf}
&\begin{array}{cccc}Q_{10}x={\psi},~~&Q_{10}z={\dot{\xi}},~~&Q_{10}\psi={\dot x},~~&Q_{10}\xi={ z},~~\\
Q_{01}x={\xi},~~&Q_{01}z=-{\dot{\psi}},~~&Q_{01}\psi=-{z},~~&Q_{01}\xi={\dot x},~~\\
~~Z x={z},~~&~~Zz=-{\ddot x},~~&~~Z \psi=-{\dot\xi},~~&~~Z \xi={\dot\psi};~~
\end{array}&
\eea
\par
~\par
{\it iii}) for the $(0,2,2)$ multiplet the $D$-module representation is
{
\footnotesize{
\bea\label{root2}
H= \left(\begin{array}{cccc}\partial_{\tau}&0&0&0\\0&\partial_{\tau}&0&0\\0&0&\partial_{\tau}&0\\ 
0&0&0&\partial_{\tau}\end{array}\right), &\quad&
Z= \left(\begin{array}{cccc}0&\partial_{\tau}&0&0\\-\partial_{\tau}&0&0&0\\0&0&0&-\partial_{\tau}
\\ 0&0&\partial_{\tau}&0\end{array}\right),\nonumber\\
Q_{10}= \left(\begin{array}{cccc}0&0&\partial_\tau&0\\0&0&0&\partial_\tau\\ 1&0&0&0\\ 
0&1&0&0\end{array}\right), &\quad&
Q_{01}= \left(\begin{array}{cccc}0&0&0&\partial_\tau\\0&0&-\partial_\tau&0\\0&-1&0&0
\\ 1&0&0&0\end{array}\right).
\eea
}
}
The corresponding transformations of the component fields are
\bea
&\begin{array}{cccc}Q_{10}x={\dot\psi},~~&Q_{10}z={\dot{\xi}},~~&Q_{10}\psi={x},~~&Q_{10}\xi={z},~~\\
Q_{01}x={\dot\xi},~~&Q_{01}z=-{\dot\psi},~~&Q_{01}\psi=-{z},~~&Q_{01}\xi={x},~~\\
~~Z x={\dot z},~~&~~Zz=-{\dot x},~~&~~Z \psi=-{\dot\xi},~~&~~Z \xi={\dot\psi}.~~
\end{array}&
\eea
It is worth noticing that the $D$-module representation of  (\ref{z2z2super}) acting on the $(0,2,2)$ multiplet can also be recovered from the $(2,2,0)$ root $D$-module representation by applying a similarity transformation. Let $g$ denotes a given generator in (\ref{root}). The corresponding generator
$g'$ acting on the $(0,2,2)$ multiplet can be expressed, in terms of the $2\times 2$ matrices $Y,I$  introduced in (\ref{splitquat}), as
\bea\label{220simtran}
g &\mapsto & g' =  (Y\otimes I)\cdot g \cdot(Y\otimes I),  \qquad {\textrm{where}}\quad (Y\otimes I)^2={\mathbb I}_4.
\eea  \par
This expression for $g'$ coincides up to a sign with the corresponding generator obtained from the ${\mathfrak D}_3$ dressing and presented in (\ref{root2}).\par
Similarly, the $D$-module representations associated with the  $(1,2,1)_{[00]}$ and $(1,2,1)_{[11]}$ multiplets are interrelated by a similarity transformation. Let ${\widetilde g}$ denotes any generator given in (\ref{12111rep}),
its associated ${\widehat g}$ operator expressed by
\bea\label{121simtran}
{\widetilde g}&\mapsto {\widehat g}=(I\otimes Y) \cdot{\widetilde g}\cdot (I\otimes Y),  \qquad {\textrm{where}}\quad (I\otimes Y)^2={\mathbb I}_4,
\eea 
coincides, up to a sign, with the corresponding generator obtained from the ${\mathfrak D}_2$ dressing and presented in (\ref{12100rep}).

\section{$D$-module representations of the ${\mathbb Z}_2\times {\mathbb Z}_2$-graded conformal superalgebra}

A ${\mathbb Z}_2\times{\mathbb Z}_2$-graded conformal superalgebra extension of the 
${\mathbb Z}_2\times{\mathbb Z}_2$-graded superalgebra (\ref{z2z2super})  is obtained by introducing the conformal
partners of the generators $H, Q_{10}, Q_{01}, Z$. The minimal conformal extension ${\cal G}_{conf}$ corresponds to a superalgebra
spanned by $10$ generators. The $6$ extra generators will be denoted as $D,U, S_{10}, S_{01}, K, W$. The (anti)commutators defining ${\cal G}_{conf}$ respect both the ${\mathbb Z}_2\times{\mathbb Z}_2$-grading $ij$ and the scaling dimension \textcolor{black}{(see Appendix {\bf A})} $s$ of the generators. Scaling dimension and ${\mathbb Z}_2\times{\mathbb Z}_2$-grading of
the ${\cal G}_{conf}$ generators are assigned according to the table
\bea\label{table}
&
\relax 
\begin{array}{|l|c|c|c|c|}\hline  {~s ~} \backslash ij&$00$&$11$&$10$&$01$ \\ \hline 
+1:&H&Z&&\\ \hline 
+\frac{1}{2}:&&&Q_{10}&Q_{01}\\  \hline 
~~ 0:&D&U&&\\ \hline 
-\frac{1}{2}:&&&S_{10}&S_{01}\\  \hline 
-1:&K&{W}&&\\ \hline
\end{array}
&
\eea 
The minimal conformal extension ${\cal G}_{conf}$ can be recovered from the (\ref{root}) $D$-module root representation  (\ref{root}) of the superalgebra (\ref{z2z2super}) by adding an extra operator $K$ (the conformal partner of
$H$), which is introduced through the position
\bea\label{kroot}
K &=& -\tau^2\partial_\tau {\mathbb I}_4- 2\tau\Lambda, \qquad \Lambda=diag(\lambda,\lambda,\lambda+\frac{1}{2},\lambda+\frac{1}{2}).
\eea
The remaining generators entering table (\ref{table}) and their (anti)commutators defining ${\cal G}_{conf}$ are recovered from repeated (anti)commutators involving the operators $Q_{10}$, $Q_{01}$ and $K$. \par
The nonvanishing ${\cal G}_{conf}$ (anti)commutators are
\bea\label{conformal220}
&\begin{array}{llll}
 \relax [H, D]=-H,&[H, U]=2Z,& [H, S_{10}]=Q_{10},& [H, S_{01}]=Q_{01},\\
\relax [H, K]=2D,&  [H, W]=-U,&[Z, D]=-Z,&[Z, U]=-2H,\\
\relax  \{Z, S_{10}\}=Q_{01},&\{Z, S_{01}\}=-Q_{10},&  [Z, K]=-U,&[Z, W]=-2D ,\\
\relax  \{Q_{10},Q_{10}\}=2H,&[Q_{10}, Q_{01}]= -2Z,&[Q_{10}, D]=-\frac{1}{2}Q_{10}, &
 \{Q_{10}, U\}=-Q_{01},\\
\relax \{Q_{10}, S_{10}\}=-2D,&[Q_{10}, S_{01}]=-U,&[Q_{10}, K]=-S_{10},&\{Q_{10}, W\}=S_{01} ,
\\
\relax  \{Q_{01}, Q_{01}\}=2H,& [Q_{01}, D]=-\frac{1}{2} Q_{01}, & \{Q_{01}, U\}=Q_{10},&[Q_{01}, S_{10}]=U
,\\
\relax \{Q_{01}, S_{01}\}=-2D,&
\relax [Q_{01}, K]=-S_{01},&\{Q_{01}, W\}=-S_{10} ,& [D, S_{10}]=-\frac{1}{2} S_{10},\\
\relax  [D, S_{01}]=-\frac{1}{2}S_{01},&[D, K]=-K,&  [D, W]=-W,&\{U, S_{10}\}=S_{01},\\
\relax \{U, S_{01}\}=-S_{10},&[U, K]=2W,& [U, W]=-2K ,& \{S_{10}, S_{10}\}=-2K,\\
\relax [S_{10}, S_{01}]=2W,&\{S_{01}, S_{01}\}=-2K.&&
\end{array}&\nonumber\\&&
\eea
The closure of the ${\cal G}_{conf}$ algebra is realized for any real value of the parameter $\lambda$ entering (\ref{kroot}). 
Therefore $\lambda\in {\mathbb R}$ is unconstrained.\par
The $D$-module representation corresponding to the $(2,2,0)$ root multiplet is given by the operators

{\footnotesize{
\bea\label{220conf}
H& =&\left(\begin{array}{cccc}\partial_{\tau}&0&0&0\\0&\partial_{\tau}&0&0\\0&0&\partial_{\tau}&0\\ 
0&0&0&\partial_{\tau}\end{array}\right), \nonumber\\
Z&=& \left(\begin{array}{cccc}0&\partial_{\tau}&0&0\\-\partial_\tau&0&0&0\\0&0&0&-\partial_{\tau}
\\ 0&0&\partial_{\tau}&0\end{array}\right),\nonumber\\
Q_{10}&=& \left(\begin{array}{cccc}0&0&1&0\\0&0&0&1\\ \partial_\tau&0&0&0\\ 
0&\partial_{\tau}&0&0\end{array}\right), \nonumber\\
Q_{01}&=& \left(\begin{array}{cccc}0&0&0&1\\0&0&-1&0\\0&-\partial_{\tau}&0&0
\\ \partial_\tau&0&0&0\end{array}\right),\nonumber\\
D& =&\left(\begin{array}{cccc}-\tau\partial_{\tau}-\lambda&0&0&0\\0&-\tau\partial_{\tau}-\lambda&0&0\\0&0&-\tau\partial_{\tau}-(\lambda+\frac{1}{2})&0\\ 
0&0&0&-\tau\partial_{\tau}-(\lambda+\frac{1}{2})\end{array}\right), \nonumber\\
{U}&=& \left(\begin{array}{cccc}0&2(\tau\partial_{\tau}+\lambda)&0&0\\-2(\tau\partial_{\tau}+\lambda)&0&0&0\\0&0&0&-(2\tau\partial_{\tau}+2\lambda+1)
\\ 0&0&2\tau\partial_{\tau}+2\lambda+1&0\end{array}\right),\nonumber\\
{S}_{10}&=& \left(\begin{array}{cccc}0&0&\tau&0\\0&0&0&\tau\\ \tau\partial_\tau+2\lambda&0&0&0\\ 
0&\tau\partial_{\tau}+2\lambda&0&0\end{array}\right), \nonumber\\
{S}_{01}&=& \left(\begin{array}{cccc}0&0&0&{\tau}\\0&0&-\tau&0\\0&-(\tau\partial_{\tau}+2\lambda)&0&0
\\ \tau\partial_\tau+2\lambda&0&0&0\end{array}\right),\nonumber\\
K& =&\left(\begin{array}{cccc}-\tau^2\partial_{\tau}-2\tau\lambda&0&0&0\\0&-\tau^2\partial_{\tau}-2\tau\lambda&0&0\\0&0&-\tau^2\partial_{\tau}-(2\lambda+1)\tau&0\\ 
0&0&0&-\tau^2\partial_{\tau}-(2\lambda+1)\tau\end{array}\right), \nonumber\\
{W}&=& \left(\begin{array}{cccc}0&-\tau^2\partial_{\tau}-2\lambda\tau&0&0\\\tau^2\partial_{\tau}+2\lambda\tau&0&0&0\\0&0&0&\tau^2\partial_{\tau}+(2\lambda+1)\tau
\\ 0&0&-\tau^2\partial_{\tau}-(2\lambda+1)\tau&0\end{array}\right).
\eea
}}

The ${\cal G}_{conf}$ algebra contains several subalgebras. In particular the $sl(2)$ subalgebra generated by
$H,D,K$ where $D$, the scaling operator, is the Cartan element. Two different $osp(1|2)$ superalgebras are recovered from the subsets of generators $\{H,D,K,Q_{10}, S_{10}\}$ and $\{H,D,K,Q_{01}, S_{01}\}$, respectively.\par

\subsection{${\mathbb Z}_2\times {\mathbb Z}_2$-graded conformal superalgebra and dressed multiplets}

The $D$-module representation of ${\cal G}_{conf}$ acting on the $(0,2,2)$ multiplet is obtained by extending the (\ref{220simtran})  similarity transformation to any generator of  ${\cal G}_{conf}$.  Let $g$ denotes a given generator in (\ref{220conf}). The corresponding generator
$g'$ acting on the $(0,2,2)$ multiplet is given by
\bea\label{similconformal022}
g &\mapsto & g' =  (Y\otimes I)\cdot g \cdot(Y\otimes I),  \qquad {\textrm{where}}\quad (Y\otimes I)^2={\mathbb I}_4,
\eea  
 with $Y$ and $I$  introduced in (\ref{splitquat}).\par

The $D$-module representation of the minimal ${\cal G}_{conf}$ algebra acting on the $(1,2,1)_{[11]}$ multiplet is obtained by applying to the (\ref{220conf}) operators the dressing transformation generated by the diagonal matrix ${\frak D}_1$ introduced in (\ref{dressingmatrices}).\par
Let $g$ be  a given operator in (\ref{220conf}), the corresponding ${\widetilde g}$ dressed operator is given by
 \bea\label{singular}
g &\mapsto & {\widetilde g}=  {\frak D}_1\cdot g \cdot{{\frak D}_1}^{-1},
\eea  
where ${\mathfrak{D}}_1$ has been introduced in (\ref{dressingmatrices}).\par
The four dressed operators ${\widetilde H}, {\widetilde Z}, {\widetilde Q}_{10}, {\widetilde Q}_{01}$ are differential
matrix operators. On the other hand, due to the presence of the inverse matrix ${{\frak D}_1}^{-1}$ in (\ref{singular}), the remaining $6$ transformed matrices ${\widetilde D}, {\widetilde U}, {\widetilde S}_{10}, {\widetilde S}_{01}, {\widetilde K}, {\widetilde W}$  are differential operators only if the real parameter $\lambda$, which is unconstrained in (\ref{220conf}),
is set to $0$:
\bea
\lambda&=&0.
\eea
Therefore, the minimal ${\cal G}_{conf}$ conformal algebra is recovered from the  $(1,2,1)_{[11]}$ multiplet by taking repeated (anti)commutators of the operators ${\widetilde Q}_{10},~ {\widetilde Q}_{01},~ {\widetilde K}_{\lambda=0}$, given by
{\footnotesize{
\bea
&&
{\widetilde Q}_{10}= \left(\begin{array}{cccc}0&0&\partial_\tau&0\\0&0&0&1\\ 1&0&0&0\\ 
0&\partial_{\tau}&0&0\end{array}\right), \qquad
{\widetilde Q}_{01}= \left(\begin{array}{cccc}0&0&0&\partial_\tau\\0&0&-1&0\\0&-\partial_{\tau}&0&0
\\ 1&0&0&0\end{array}\right),\nonumber\\
&&~~~{\widetilde K}_{\lambda=0}=\left(\begin{array}{cccc}-\tau^2\partial_{\tau}-2\tau&0&0&0\\0&-\tau^2\partial_{\tau}&0&0\\0&0&-\tau^2\partial_{\tau}-\tau&0\\ 
0&0&0&-\tau^2\partial_{\tau}-\tau\end{array}\right).
\eea
}}
A nonminimal ${\mathbb Z}_2\times{\mathbb Z}_2$-graded conformal extension of ${\cal G}_{conf}$, requiring the introduction of new generators, is recovered by taking repeated (anti)commutators of the operators ${\widetilde Q}_{10}, ~{\widetilde Q}_{01}$ and $ {\widetilde K}_{\lambda}$, where $ {\widetilde K}_{\lambda}$ is defined for $\lambda\neq 0$ as
\bea
{\widetilde K}_{\lambda}&=& {\widetilde K}_{\lambda=0}-2\lambda\tau\cdot{\mathbb I}_4.
\eea
This new nonminimal algebra is denoted as ${\cal G}_{nm,conf}$. \par

{\textcolor{black}{
${\cal G}_{nm,conf}$ is finitely generated by the the three generators ${\widetilde Q}_{10}$, ${\widetilde Q}_{01}$, ${\widetilde K}_\lambda$. Each generator entering ${\cal G}_{nm,conf}$ is obtained by taking repeated
(anti)commutators involving ${\widetilde Q}_{10}$, ${\widetilde Q}_{01}$, ${\widetilde K}_\lambda$. For instance,
${\widetilde H}$ is recovered from the anticommutator $\{{\widetilde Q}_{10},{\widetilde Q}_{10}\}=2{\widetilde H}$,  
${\widetilde Z}$ from the commutator $[{\widetilde Q}_{10},{\widetilde Q}_{01}]=-2{\widetilde Z}$, while
${\widetilde U}$, belonging to the $11$-sector, from $[{\widetilde Z},{\widetilde K}]=-2{\widetilde U}$ and so on.}}
\par
{\textcolor{black}{For $\lambda\neq 0$ ${\cal G}_{nm,conf}$ is an infinitely dimensional ${\mathbb Z}_2 \times {\mathbb Z}_2$-graded Lie superalgebra, possessing an infinite number of generators.  This is seen as follows: the commutator between the $11$-graded generators ${\widetilde Z}, {\widetilde U}$ produces
$[{\widetilde Z},{\widetilde U}]= -2{\widetilde H}-4\lambda {\widetilde H}_2$, where ${\widetilde H}_2$ is a new $00$-graded generator, given by {\footnotesize {${\widetilde H}_2 = \left(\begin{array}{cccc}-\partial&0&0&0\\0&\partial&0&0\\ 0&0&0&0
\\0&0&0&0 \end{array}\right)$.}}
The commutator 
$[{\widetilde Z},{\widetilde H}_2]= 2{\widetilde Z}_2$ produces the new $11$-graded generator {\footnotesize{${\widetilde Z}_2 = \left(\begin{array}{cccc}0&\partial^3&0&0\\\partial&0&0&0\\ 0&0&0&0
\\0&0&0&0 \end{array}\right)$}}.
By taking repeated commutators with ${\widetilde Z}$ one generates an infinite tower of new generators ${\widetilde H}_n$, ${\widetilde Z}_n$. This shows that, for $\lambda\neq 0$, ${\cal G}_{nm,conf}$ is an infinite-dimensional  
${\mathbb Z}_2 \times {\mathbb Z}_2$-graded Lie superalgebra.\\
The generator {\footnotesize{${\widetilde M}=\left(\begin{array}{cccc}0&0&0&0\\0&0&0&0\\ 0&0&0&1
\\0&0&1&0 \end{array}\right)$}}, obtained from the right hand side of
\bea\label{nonvanish}
\relax [{\widetilde Q}_{10}, {\widetilde S}_{01}] +[{\widetilde Q}_{01}, {\widetilde S}_{10}] &=&4\lambda {\widetilde M},
\eea
 is another extra generator entering ${\cal G}_{nm,conf}$.}}\par

We finally mention that  the $D$-module representations of the  conformal algebras associated with the  $(1,2,1)_{[00]}$ multiplet are recovered from the $(1,2,1)_{[11]}$ representations by applying an extension of the  (\ref{121simtran}) similarity transformation. Let ${\widetilde g}$ denotes any conformal generator associated with the $(1,2,1)_{[11]}$ multiplet,
the corresponding ${\widehat g}$ generator associated with the $(1,2,1)_{[00]}$ multiplet is given by
\bea
{\widetilde g}&\mapsto {\widehat g}=(I\otimes Y) \cdot{\widetilde g}\cdot (I\otimes Y),  \qquad {\textrm{where}}\quad (I\otimes Y)^2={\mathbb I}_4,
\eea 
 for $Y$, $I$ introduced in (\ref{splitquat}).

\section{${\mathbb Z}_2\times {\mathbb Z}_2$-graded invariant actions}

In this Section we present a general framework to construct ${\mathbb Z}_2\times {\mathbb Z}_2$-graded classical invariant actions, in the Lagrangian setting, for the basic multiplets introduced in Section {\bf 3}. The approach works for both single basic multiplets and for several interacting basic multiplets. We discuss, at first, the actions for the root multiplet. The modifications to be applied for the construction of actions for the dressed multiplets are immediate.

\subsection{Invariant actions for the  $(2,2,0)$ root multiplet}

We rely on the fact that the differential operators introduced in (\ref{root}) satisfy the ${\mathbb Z}_2\times {\mathbb Z}_2$-graded Leibniz rule when acting on functions of the component fields $x,z,\psi,\xi$. The actions and the Lagrangians are required to belong to the $00$-graded sector. \par
Therefore, a manifestly invariant sigma-model action ${\cal S}_\sigma$ for the $(2,2,0)$ multiplet can be expressed as
\bea\label{sigma220a}
{\cal S}_\sigma&=& \int d\tau {\cal L_\sigma}, \quad\quad {\cal L}_\sigma= ZQ_{10}Q_{01}g(x,w), \quad{\textrm{for}} \quad w=z^2.
\eea
In the above formula $g(x,w)$ is an arbitrary $00$-graded prepotential of the even fields $x,z$. Due to the (\ref{z2z2super}) (anti)commutators and their explicit expression, the action of $H,Z, Q_{10}, Q_{01}$ on ${\cal L}_\sigma$ produces
a time derivative, making the ${\cal S}_\sigma$ action ${\mathbb Z}_2\times {\mathbb Z}_2$-graded invariant. Up to boundary terms we  have
\bea\label{sigmaphi}
{\cal L}_\sigma &\sim& \Phi(x,w)({\dot x}^2+{\dot z}^2-\psi{\dot\psi}+\xi{\dot \xi})+(\Phi_x{\dot z}-2\Phi_w {\dot x}z)\psi\xi,
\eea
where
\bea
\Phi(x,w) &=& g_{xx}+2g_w+4wg_{ww}
\eea
(here and in the following the suffix denotes derivative with respect to the corresponding field so that, e.g., 
$\Phi_x(x,w) = \textcolor{black}{\frac{\partial \Phi(x,w)}{\partial x}}$).\par
The invariant sigma-model defined by (\ref{sigma220a}) is not the most general one. Another 
manifestly invariant $(2,2,0)$ action ${\cal S}_{\overline\sigma}$ is obtained from setting
\bea\label{sigma220b}
{\cal S}_{\overline\sigma}&=&\int d\tau ZQ_{10}Q_{01} \left(f(x,w) z\psi\xi\right),
\eea
where the new prepotential $f(x,w) z\psi\xi$ also belongs to the $00$-graded sector. Since the odd-fields $\psi,\xi$ are Grassmann, the most general manifestly invariant sigma-model is produced by the linear combination 
${\cal S}= {\cal S}_{\sigma}+{\cal S}_{\overline\sigma}$ for arbitrary prepotentials $ g(x,w),~f(x,w) z\psi\xi$.\par
One should warn that, contrary to $g(x,w)$, the $f(x,w) z\psi\xi$ prepotential in (\ref{sigma220b}) produces higher derivatives. Indeed the simplest choice, obtained by setting $f(x,w)=1$, produces the Lagrangian
\bea
{\overline{\cal L}} &=& ZQ_{10}Q_{01}(z\psi\xi) \sim Z(z({\dot x}^2+{\dot z}^2-\psi{\dot\psi}+\xi{\dot \xi})-{\dot x}\psi\xi)
\eea
which contains  a third order time derivative in $x$ (${\overline{\cal L}}={\dot x}^3+\ldots$) that cannot be reabsorbed by a total time derivative.\par
The free kinetic action is defined by setting $\Phi(x,w)=\frac{1}{2}$ in (\ref{sigmaphi}). The corresponding Lagrangian ${\cal L}$, given by
\bea
{\cal L}&=&   \frac{1}{2}({\dot x}^2+{\dot z}^2-\psi{\dot\psi}+\xi{\dot \xi}),
\eea
is invariant under the  full (\ref{conformal220})  ${\mathbb Z}_2\times {\mathbb Z}_2$-graded conformal superalgebra ${\cal G}_{conf}$. This is a consequence of the relation
\bea
K{\cal L} &=& \frac{1}{2}\frac{d}{d\tau} \left(-\tau^2{\dot x}^2+x^2-\tau^2{\dot z}^2+z^2 +\tau^2\psi{\dot\psi}-\tau^2\xi{\dot\xi}\right),
\eea
for $K$ given by (\ref{220conf}) with the 
\bea
\Lambda &=&diag (-\frac{1}{2}, -\frac{1}{2}, 0,0)
\eea
assignment of the scaling dimensions of the root multiplet component fields.\par
We can generalize the manifestly invariant action (\ref{sigma220a}) to the case of $n$ independent root multiplets
labeled by $i=1,2,\ldots, n$. In each multiplet its component fields 
$x_i, \psi_i,\xi_i, z_i$ transform according to (\ref{rootfieldtransf}). An invariant sigma-model action ${\cal S}_{int}$, describing the motion of interacting multiplets,
can be defined through the position
\bea\label{sigmainteract}
{\cal S}_{int}&=& \int d\tau ZQ_{10}Q_{01}g(x_i,w_{ij}), 
\eea
for a generic prepotential $g(x_i, w_{ij})$, with $w_{ij}=z_iz_j$. The introduction of nontrivial interactions among multiplets requires suitably choosing the prepotential function $g(x_i,w_{ij})$.  One can set, e.g., $\partial_{x_k}\partial_{x_l}g(x_i,w_{ij})$ for $k\neq l$ to be nonvanishing functions of the component fields.

\subsection{Invariant actions for the $(1,2,1)_{[00]}$ multiplet}

As the next case we are considering the invariant actions for the dressed $(1,2,1)_{[00]}$ multiplet with an ordinary propagating boson. Its component fields $x, z,\psi,\xi$ transform according to \textcolor{black}{  \eqref{12100transf}.}
\par
A sigma-model type of action, the counterpart of (\ref{sigmaphi}), can be formally expressed with the same notation,
but taking into account the different role of the exotic boson $z$:
\bea\label{sigma12100}
{\cal S}_\sigma&=& \int d\tau {\cal L_\sigma}, \quad\quad {\cal L}_\sigma= ZQ_{10}Q_{01}g(x,w), \quad{\textrm{for}} \quad w=z^2.
\eea
Up to boundary terms the Lagrangian ${\cal L}_\sigma$ now reads
\bea
{\cal L_\sigma}&\sim& (g_{xxx}-2g_{xxw}{\ddot x}) z\psi\xi+[(2 g_{xxw}w+g_{xx})-2h_x{\ddot x}](\xi{\dot \xi}-\psi{\dot\psi})+
2h ({\dot\xi}{\ddot \xi}-{\dot\psi}{\ddot\psi})+\nonumber\\
&&(4g_{xw}+2h_x-4h_w{\ddot x})z{\dot\psi}{\dot\xi}+g_{xx}\textcolor{black}{ w}-(4g_{xw}w+g_x){\ddot x}+\nonumber\\
&&{\textcolor{black}{2h{\ddot x}^2-2g_wz{\ddot z}-2g_{xw}z({\ddot\psi}\xi+\psi{\ddot\xi}),}}
\eea
where $ h(x,w)$ is introduced as
\bea
h(x,w)&=&2g_{ww}w+g_w.
\eea
Let us present several particular cases:
\begin{enumerate}
\item [1)] for $g(x,w) = g(x)$ we have
\bea\label{sigma121special}
{\cal L}_\sigma&\sim& g_{xx}({\dot x}^2+z^2-\psi{\dot\psi}+\xi{\dot \xi})+g_{xxx}z\psi\xi;
\eea
\item[2)] for $g(x,w)=g(w)$ we have $h(x,w)=h(w)$. The Lagrangian ${\cal L}_\sigma$  possesses higher-order time derivatives,
\bea
{\cal L}_\sigma&\sim& 2h({\ddot x}^2+{\dot z}^2-{\dot\psi}{\ddot\psi}+{\dot\xi}{\ddot\xi})-4h_w{\ddot x}z{\dot\psi}{\dot\xi};
\eea
\item[3)] the condition $h(x,w)=0$ implies that the prepotential $g(x,w)$ has the form $$g(x,w)=a(x)\sqrt{w}+b(x).$$ Under this condition the Lagrangian ${\cal L}_\sigma$ admits up to a second order time derivative. The result of the computation of the second term, $b(x)$, is recovered from the result at item $1$.  The new contribution for $a(x)\neq 0$, $b(x)=0$ reads
\bea
{\cal L}_\sigma&\sim& a_{xx}\sqrt{w}(3{\dot x}^2+z^2-2\psi{\dot\psi}+2\xi{\dot\xi})+2a_x{\sqrt w}z{\dot\psi}{\dot\xi}+4\frac{a_x}{\sqrt w}z{\dot x}{\dot z}+\nonumber\\
&&\textcolor{black}{(a_{xxx}{\sqrt w}-\frac{a_{xx}}{\sqrt w}{\ddot x})z\psi\xi}
 \textcolor{black}{  - \frac{a_x}{\sqrt{w}} z (\ddot{\psi} \xi + \psi \ddot{\xi}).}
\eea 
\end{enumerate}

Let us focus our discussion on the sigma-model Lagrangian (\ref{sigma121special}). 
By setting 
$\phi(x)= 2g_{xx}$ {\textcolor{black}{it differs from the ${\cal N}=2$ supersymmetric Lagrangian (also denoted as 
${\cal L}_\sigma$)  entering formula (\ref{totallag}), in the sign in front of the $\xi{\dot \xi}$ term.}}
{\textcolor{black}{This difference is due to the ${\mathbb Z}_2\times{\mathbb Z}_2$ grading;  in particular the (\ref{sigma121special}) operators
$Q_{10}$, $Q_{01}$, unlike the $Q_1,Q_2$ supersymmetry operators used in the derivation of  (\ref{totallag}),  act on the ${\mathbb Z}_2\times{\mathbb Z}_2$-graded component fields as ${\mathbb Z}_2\times{\mathbb Z}_2$-graded Leibniz derivatives.}}
\par
{\textcolor{black}{It is worth mentioning that the sigma-model action (57), when specialized to the choice $g_{xx}= x^\alpha$ for $\alpha\neq -2$, is invariant under the nonminimal ${\mathbb Z}_2\times{\mathbb Z}_2$ conformal algebra ${\cal G}_{nm, conf}$ with the identification $\lambda= -\frac{1}{\alpha+2}$. 
}}
\par
Since $\psi, \xi$ are classical Grassmann fields (satisfying, in particular, $\psi^2=\xi^2=0$), 
\textcolor{black}{by solving the algebraic equation of motion for $z$,} 
the new Lagrangian reads
\bea\label{sigmaof121}
{\cal L}_\sigma&\sim& \frac{1}{2}\phi(x) ({\dot x}^2 -\psi{\dot\psi}+\xi{\dot\xi}).
\eea
By setting, as in formula (\ref{barredfieldsapp}),
\bea\label{cxnew}
&{\overline x}= C(x),\quad
{\overline\psi} = C_x\psi,\quad 
{\overline \xi}= C_x \xi,& \qquad {\textrm{where}}\quad
C_x= \sqrt{\phi},
\eea
we realize that the Lagrangian (\ref{sigmaof121}) corresponds to the non-interacting constant kinetic Lagrangian
${\cal L}_{kin}$ for the barred fields
\bea
{\cal L}_{kin}&=& \frac{1}{2} ({\dot {\overline x}}^2 -{\overline \psi}{\dot{\overline \psi}}+{\overline \xi}{\dot{\overline \xi}}).
\eea
The  introduction of interacting terms is reached by adding, as also discussed in the Appendix, a linear potential term in $z$ to the
sigma-model Lagrangian (\ref{sigma121special}). Since $z$ transforms as a time-derivative under the (\ref{12100rep}) operators, the total action ${\cal S}$ is invariant. This action is
\bea
{\cal S} &=& \int d\tau \frac{1}{2}\left(\phi(x)({\dot x}^2+z^2-\psi{\dot\psi}+\xi{\dot \xi})+\phi_xz\psi\xi+\mu z\right).
\eea
{\textcolor{black}{
In order for the action to be $00$-graded, the constant parameter $\mu$ should be $11$-graded. We recall that
the classical ${\mathbb Z}_2\times {\mathbb Z}_2$-graded fields (anti)commute, satisfying the (anti)commutators (\ref{anticommrel}).
This is a consistent extension of ordinary classical supermechanics whose fields are assumed to be real or complex (the bosons) and Grassmann (the fermions). We have now two types of Grassmann fields ($10$-graded and $01$-graded) and two types of bosons ($00$-graded and $11$-graded). They are time-dependent and their ordering is determined
by the ${\mathbb Z}_2\times {\mathbb Z}_2$-graded structure. E.g.,  following the (\ref{anticommrel}) prescription, if $z(\tau)$ is $11$-graded and $\psi(\tau)$ is $10$-graded, then
$z(\tau)\psi(\tau)=-\psi(\tau)z(\tau)$. 
The same prescription (\ref{anticommrel}) holds for ${\mathbb Z}_2\times {\mathbb Z}_2$-graded constant (not depending on $\tau$) fields. 
The $11$-graded coupling constant $\mu$ is such an example. It can be interpreted as a non-dynamical, constant, $11$-graded, background field. When quantizing the theory as discussed in \cite{akt}, $\mu$ becomes a constant $4\times 4$ matrix belonging to the ${\cal G}_{11}$ sector in formula (\ref{gradedmatrices}). At a classical level $\mu$ is assumed to (anti)commute
with the ${\mathbb Z}_2\times{\mathbb Z}_2$-graded fields. 
}}\par
By repeating the computations for the analogous case presented in the Appendix,  one can solve the algebraic equation of motion for $z$ so that, up to boundary terms, the Lagrangian ${\cal L}$ can be expressed as
\bea\label{intermediate11}
{\cal L}&=& \frac{1}{2}({\dot {\overline x}}^2-{\overline\psi}{\dot{\overline\psi}}+{\overline \xi}{\dot{\overline\xi}})
-\frac{1}{8}
(\frac{\mu}{C_x})^2-\frac{\mu}{2}\frac{C_{xx}}{(C_x)^3} {\overline \psi}{\overline \xi},
\eea
where the barred fields and  $C(x)$ are given in (\ref{cxnew}). After setting
\bea\label{Wpot}
&W({\overline x})=W(C(x))=\frac{\mu}{2C_x(x)}&
\eea
we can rewrite the Lagrangian as
\bea\label{the12100model}
{\cal L}&=& \frac{1}{2}({\dot {\overline x}}^2-{\overline\psi}{\dot{\overline\psi}}+{\overline \xi}{\dot{\overline\xi}})-\frac{1}{2}W^2({\overline x})+W_{\overline x}{\overline\psi}{\overline \xi}.
\eea
{\textcolor{black}{
The potential $W({\overline x})$ in (\ref{Wpot}) is proportional to $\mu$ and is $11$-graded since $C(x)$, defined in (\ref{cxnew}), is $00$-graded. The Lagrangians (\ref{intermediate11}) and (\ref{the12100model}) are $00$-graded.  In particular, the cubic term $\mu{\overline\psi}{\overline \xi}$ entering the right hand side of (\ref{intermediate11}) is $00$-graded since, from (\ref{anticommrel}) and (\ref{scalarproduct}), $[00] = [11]+[10]+[01]$, where the addition is $mod~2$. The consistency of the procedure follows from respecting the ${\mathbb Z}_2\times {\mathbb Z}_2$-graded properties.}}\par
This construction of the interacting action follows the second approach described in the Appendix. The first approach which works nicely for the ${\cal N}=2$ supersymmetric action and is based on a constant kinetic term plus a potential term, cannot be repeated in the ${\mathbb Z}_2\times{\mathbb Z}_2$-graded case. The reason is that the only ${\mathbb Z}_2\times{\mathbb Z}_2$ invariant potential term is given by the linear term in $z$. To get the nontrivial interaction one is therefore obliged to add this linear term to the sigma-model action and perform the (\ref{cxnew}) field redefinitions.

\subsection{Interacting $(1,2,1)_{[00]}$ multiplets }

The construction of invariant actions for interacting multiplets proceeds as for the root multiplets case. We explicitly present it for two interacting multiplets.
The fields are denoted as $x_1, z_1, \psi_1,\xi_1$ and $x_2,z_2,\psi_2,\xi_2$, respectively.
They transform independently; nevertheless, their interaction can be induced by the prepotential.\par
The sigma-model action can be defined as before, so that 
\bea
{\cal S}_\sigma=\int d\tau {\cal L}_\sigma= \int d\tau ZQ_{10}Q_{01} g(x_1,x_2).
\eea
Up to boundary terms, the Lagrangian is
\bea
{\cal L}_{\sigma}&=& g_{11}({\dot x}_1^2+z_1^2-\psi_1{\dot\psi}_1+\xi_1{\dot\xi}_1)+g_{22}({\dot x}_2^2+z_2^2-\psi_2{\dot\psi}_2+\xi_2{\dot\xi}_2)+\nonumber\\
&&+g_{12}(2{\dot x}_1{\dot x}_2+2z_1z_2-\psi_1{\dot\psi}_2-\psi_2{\dot\psi}_1+\xi_1{\dot \xi}_2+\xi_2{\dot \xi}_1)+g_{111}z_1\psi_1\xi_1+g_{222}z_2\psi_2\xi_2+\nonumber\\&&+g_{112}(z_2\psi_1\xi_1+z_1(\psi_1\xi_2+\psi_2\xi_1))+
g_{221}(z_1\psi_2\xi_2+z_2(\psi_1\xi_2+\psi_2\xi_1))
\eea
\textcolor{black}{
where $ g_{12} := \partial_{x_1} \partial_{x_2} g(x_1,x_2),$ etc.}
}
The necessary condition $g_{12}\neq 0$ is required to have interacting multiplets.\par 
The action ${\cal S}$, obtained by adding the linear potential term, is also invariant:
\bea
{\cal S}&=& \int d\tau\left( {\cal L}_\sigma +{\cal L}_{lin}\right), \qquad {\textrm{where}}\quad {\cal L}_{lin}= \mu_1z_1+\mu_2z_2,
\eea

{\textcolor{black}{For consistency, the $\mu_{1,2}$ constants belong to the $11$-graded sector; their (anti)commutation properties with respect to the ${\mathbb Z}_2\times{\mathbb Z}_2$-graded fields are defined in accordance with this position.}}\par
Extending this construction to the case of  $n>2$ interacting multiplets is immediate.  

\subsection{Invariant actions for the $(1,2,1)_{[11]}$ multiplet}

The sigma-model invariant action is expressed as
\bea
{\cal S}&=& \int d\tau {\cal L}_\sigma = \int d\tau ZQ_{10}Q_{01} f(z),
\eea
where $f(z)$ is an even function of $z$.
The computation produces, up to boundary term,
\bea
{\cal L}_\sigma&\sim& \Phi(z) ({\dot z}^2+x^2-\psi{\dot\psi}+\xi{\dot\xi})-\textcolor{black}{\Phi_z(z)}x\psi\xi,
\eea
where
\bea
\Phi(z)&=& f_{zz}(z)
\eea
is also an even function of $z$.
\par
Just like the $(1,2,1)_{[00]}$ case, an invariant linear potential term ${\cal L}_{lin}$  can be added. For this multiplet the total Lagrangian ${\cal L}$ is
\bea
{\cal L}&=& \Phi(z) ({\dot z}^2+x^2-\psi{\dot\psi}+\xi{\dot\xi})-\Phi_zx\psi\xi + \mu x,
\eea
where $\mu$ is an ordinary real (i.e., not exotic) coupling constant.

\subsection{Invariant action of the $(0,2,2)$ multiplet}

The free kinetic Lagrangian
{\textcolor{black}{\bea
{\cal L}&=&   \frac{1}{2}({x}^2-{z}^2-\psi{\dot\psi}-\xi{\dot \xi})
\eea
}}
defines the invariant action ${\cal S}=\int d\tau {\cal L}$ of the $(0,2,2)$ multiplet. The scaling dimension of the fields is
$\lambda=\frac{1}{2}$ for the even fields $x$, $z$  and  $\lambda =0$ for the odd fields $\psi$, $\xi$.\par
With respect to the differential operators  defined in (\ref{similconformal022}), the action ${\cal S}$ is invariant under the
$10$-generator ${\mathbb Z}_2\times {\mathbb Z}_2$-graded ${\cal G}_{conf}$ conformal superalgebra (\ref{conformal220}).

\section{Conclusions}
\textcolor{black}{
In the supersymmetric literature the term ``supermechanics" refers to classical systems formulated in the Lagrangian setting.} 
{\textcolor{black}{There are hundreds, possibly thousands, of papers devoted on this topic. Somewhat surprisingly, after more than fifty years since the introduction (inspired by superalgebras) of ${\mathbb Z}_2\times{\mathbb Z}_2$-graded superalgebras \cite{{rw1},{rw2},{sch}}, no work has been presented yet to analyze ${\mathbb Z}_2\times{\mathbb Z}_2$-graded symmetries in this context. To fill this vacuum  is the main motivation of the present paper. }}\par
\textcolor{black}{Our basic strategy is to mimick, as much as possible, the construction of supermechanics based on supermultiplets and their derived invariant actions. We extended to the ${\mathbb Z}_2\times{\mathbb Z}_2$-graded case the approaches of \cite{{PaTo},{KRT}} for the one-dimensional super-Poincar\'e algebras and \cite{KT} for the superconformal algebras.}\par
\textcolor{black}{We derived the basic ${\mathbb Z}_2\times{\mathbb Z}_2$-graded multiplets (even in the conformal case) and presented a general framework to construct the actions.
As a consequence, a plethora of ${\mathbb Z}_2\times{\mathbb Z}_2$-graded invariant actions has been obtained
(for single basic multiplets, for interacting multiplets, for systems with or without higher derivatives, etc.). The simplest models with ${\mathbb Z}_2\times {\mathbb Z}_2$-graded conformal invariance have also been presented.}\par
\textcolor{black}{The ${\mathbb Z}_2\times{\mathbb Z}_2$-graded invariance poses further restrictions, with respect to ordinary superalgebras, on the invariant actions and the procedures to obtain them. As an example, only the second approach described in Appendix {\bf B} for the ${\cal N}=2$ supersymmetric model can be applied to derive its $
{\mathbb Z}_2\times{\mathbb Z}_2$ counterpart given in (\ref{the12100model}).}\par
\textcolor{black}{As already mentioned in the Introduction, there has been recently a renewal of interest, which started from the works \cite{{aktt1}, {aktt2}} and \cite{{Bru},{BruDup},{NaAmaDoi},NaAmaDoi2}, in analyzing ${\mathbb Z}_2\times{\mathbb Z}_2$-graded symmetries in the context of dynamical systems. The present paper fits into this current trend. }\par
\textcolor{black}{Several open questions have yet to be answered. The most relevant ones are perhaps ``which is the quantum role of the $11$-graded exotic bosons?" and ``which is the quantum signature of a ${\mathbb Z}_2\times{\mathbb Z}_2$-graded symmetry?".  }
\par
\textcolor{black}{On an abstract level the ${\mathbb Z}_2\times{\mathbb Z}_2$-graded symmetry is related with a specific form of parastatistics, see \cite{tol,StVdJ}. Concretely, the existence of a ${\mathbb Z}_2\times {\mathbb Z}_2$-symmetry tells us that multiparticle
states can be (anti)symmetrized according to the ${\mathbb Z}_2\times{\mathbb Z}_2$-statistics.  These states do not obey the
ordinary boson-fermion statistics. The use of the alternative ${\mathbb Z}_2\times {\mathbb Z}_2$ statistics has consequences which are measurable and can be observed. It affects, e.g., the
energy degeneracy of multiparticle wavefunctions, the partition function, the derived chemical potentials, etc.}
\par

\par

~\par
~\par
\par {\Large{\bf Acknowledgments}}
{}~\par{}~\par

Z. K. and F. T. are grateful to the Osaka Prefecture University, where this work was completed, for hospitality.
 F. T. was supported by CNPq (PQ grant 308095/2017-0). 

\par

~\par
~\par

  \renewcommand{\theequation}{A.\arabic{equation}}
  \setcounter{equation}{0}  

\textcolor{black}{
{\Large{\bf{Appendix A: the scaling dimensions}}} }\par
~\par

Besides the ${\mathbb Z}_2\times {\mathbb Z}_2$-grading, a scaling dimension can be assigned to the component fields entering the (\ref{fourmultiplets}) multiplets. Let us assign to the Euclidean time $\tau$ the scaling dimension 
\bea
[\tau]&=&-1.
\eea
By consistency, the scaling dimension of the ${\mathbb Z}_2\times {\mathbb Z}_2$-graded superalgebra
operators entering  (\ref{z2z2super}) are
\bea
[H]=[Z]= 1, \quad &&\quad [Q_{10}]=[Q_{01}]= \frac{1}{2}.
\eea
For each multiplet a scaling dimension can be assigned to its component fields in terms of an arbitrary real parameter $\lambda\in{\mathbb R}$. The parameter $\lambda$, which coincides  the lowest scaling dimension of a component field in a given multiplet, is called the scaling dimension of the multiplet. \par
~\par
The consistent assignment of scaling dimensions are\par
~\par
{\it i}) for the $(2,2,0)$ root multiplet,
\bea
&[x] = [z]=\lambda, \quad \quad  [\psi]=[\xi]=\lambda+\frac{1}{2};&
\eea

{\it ii}) for the $(1,2,1)_{[11]}$ ${\mathbb Z}_2\times {\mathbb Z}_2$-graded multiplet,
\bea
&[x] = \lambda +1, \quad\quad [z]=\lambda,  \quad \quad [\psi]=[\xi]=\lambda+\frac{1}{2}.&
\eea

{\it iii})
for the $(1,2,1)_{[00]}$ ${\mathbb Z}_2\times {\mathbb Z}_2$-graded multiplet,
\bea
&[x] = \lambda, \quad\quad [z]=\lambda+1,  \quad \quad [\psi]=[\xi]=\lambda+\frac{1}{2};&
\eea

{\it iv}) for the $(0,2,2)$ ${\mathbb Z}_2\times {\mathbb Z}_2$-graded multiplet,
\bea
&[x] = [z]=\lambda+\frac{1}{2}, \quad \quad  [\psi]=[\xi]=\lambda.&
\eea
\par
~\par
Let us set
\bea
&\lambda_1=[x],\quad \lambda_2=[z],\quad \lambda_3=[\psi],\quad \lambda_4=[\xi].&
\eea
For each one of the four cases above a scaling operator $D$ defines the scaling dimension of the operators
$H, Z, Q_{10}, Q_{01}$. The scaling dimension is read from the commutators
\bea
&[D,H] =H, \quad [D,Z]= Z, \quad [D, Q_{10}]=\frac{1}{2} Q_{10}, \quad [D, Q_{01}]=\frac{1}{2} Q_{01}.
\eea
The operator $D$ can be introduced through the position
\bea\label{scalingop}
D&=& -\tau\partial_\tau\cdot {\mathbb I}_4 -\Lambda, \quad {\textrm{for}}\quad  \Lambda=diag(\lambda_1,\lambda_2,\lambda_3,\lambda_4).
\eea
In the above formula $\Lambda$ is a diagonal operator. One should note that $D$ is an operator belonging to the ${\cal G}_{00}$ sector of the ${\mathbb Z}_2\times {\mathbb Z}_2$-graded superalgebra.\par
For several applications it is important to mention that constant matrices $M$, possessing a non-vanishing scaling dimension as  defined by $D$, exist in each one of the different $D$-module representations. The scaling dimension $s$
is given by
\bea
[D,M]&=&s M.
\eea
Let $E_{ij}$ denotes the matrix with entry $1$ at the intersection of the $i$-th column with the $j$-th row and $0$ otherwise. The constant matrices with non-vanishing scaling dimensions are:\par
~\par
{{
{\it i}) for the $(2,2,0)$ $D$-module representation,
\bea
s=\frac{1}{2}:&& {\textrm{for}} \quad E_{13}, E_{24}\in {\cal G}_{10}\quad {\textrm{and}}\quad E_{14}, E_{23}\in {\cal G}_{01},\nonumber\\
s=-\frac{1}{2}:&& {\textrm{for}} \quad E_{31}, E_{42}\in {\cal G}_{10}\quad {\textrm{and}}\quad E_{32}, E_{41}\in {\cal G}_{01};
\eea

{\it ii}) for the $(1,2,1)_{[11]}$ $D$-module representation,
\bea
s=1:&& {\textrm{for}} \quad E_{21}\in {\cal G}_{11},\nonumber\\
s=\frac{1}{2}:&& {\textrm{for}} \quad E_{24}, E_{31}\in {\cal G}_{10}\quad {\textrm{and}}\quad E_{23}, E_{41}\in {\cal G}_{01},\nonumber\\
s=-\frac{1}{2}:&& {\textrm{for}} \quad E_{13, E_{42}}\in {\cal G}_{10}\quad {\textrm{and}}\quad E_{14}, E_{32}\in {\cal G}_{01},\nonumber\\
s=-1:&& {\textrm{for}} \quad E_{12}\in {\cal G}_{11};
\eea

{\it iii}) for the $(1,2,1)_{[00]}$ $D$-module representation,
\bea
s=1:&& {\textrm{for}} \quad E_{12}\in {\cal G}_{11},\nonumber\\
s=\frac{1}{2}:&& {\textrm{for}} \quad E_{13}, E_{42}\in {\cal G}_{10}\quad {\textrm{and}}\quad E_{14}, E_{32}\in {\cal G}_{01},\nonumber\\
s=-\frac{1}{2}:&& {\textrm{for}} \quad E_{24}, E_{31}\in {\cal G}_{10}\quad {\textrm{and}}\quad E_{23}, E_{41}\in {\cal G}_{01},\nonumber\\
s=-1:&& {\textrm{for}} \quad E_{21}\in {\cal G}_{11}.
\eea
}}

\par
~\par
  \renewcommand{\theequation}{B.\arabic{equation}}
  \setcounter{equation}{0}  

{\Large{\bf{Appendix B: revisiting the ${\cal N}=2$ supersymmetric 
action \\ $~~~~~~~~~~~~~~~~~~~~~$ for the  real supermultiplet }}}

\par
~\par

The ${\cal N}=2$ supersymmetric action of the real supermultiplet is well known {\textcolor{black}{ \cite{{DVF},{witten},{FreTow}}.}} It consists of a constant kinetic term plus a generic potential term; it is obtained either from a superfield \textcolor{black}{\cite{BellKri}} or from a $(1,2,1)$ $D$-module approach. We present here {\textcolor{black}{for illustrative purposes}} a derivation of this action as recovered from a sigma model Lagrangian 
{\textcolor{black}{plus a Fayet-Iliopoulos \cite{FayIli} linear potential term}}. {\textcolor{black}{ It is the same method which was used in Section {\bf 5} to obtain the
non-trivial ${\mathbb Z}_2\times{\mathbb Z}_2$-graded invariant actions for the $(1,2,1)_{[11]}$ and $(1,2,1)_{[00]}$ multiplets. }} {\textcolor{black}{Indeed, the ${\mathbb Z}_2\times{\mathbb Z}_2$-graded symmetry differs and is more stringent than ordinary supersymmetry, so that not all methods which work in the supersymmetric case have a counterpart which is applicable to the ${\mathbb Z}_2\times{\mathbb Z}_2$-graded case. In particular the 
${\mathbb Z}_2\times{\mathbb Z}_2$-graded symmetry forces the potential term to be linear.}}
\par
The four time-dependent fields of the ${\cal N}=2$ model are denoted as $x$ (the propagating boson), $\psi $, $\xi$ (the fermionic fields) and $z$ (the auxiliary bosonic field). Their field transformations are
\bea
&\begin{array}{llll}
Q_1x = \psi,\quad &  Q_1 z = {\dot \xi},\quad & Q_1\psi = {\dot x}, \quad& Q_1\xi = z,\\
Q_2x = \xi,\quad & Q_2 z = -{\dot \psi},\quad & Q_2\psi = -z,\quad& Q_2\xi = {\dot x}.
\end{array}&
\eea
The one-dimensional ${\cal N}=2$ supersymmetry algebra (with generators $Q_1, Q_2, H$) satisfies
\bea
\{Q_i,Q_j\}=2\delta_{ij}H, &\quad & [H,Q_i]=0, \qquad {\textrm{for}}\quad i,j=1,2.
\eea 
The standard construction of the invariant action is made through the position
\bea
{\cal S} = \int d\tau {\cal L},&&  {\textrm{where}}\quad {\cal L} = {\cal L}_{kin} + {\cal L}_{pot}.
\eea
The kinetic and potential terms of the Lagrangian are
\bea
{\cal L}_{kin} =\frac{1}{2}\left( {\dot x}^2+z^2 -\psi{\dot\psi} - \xi{\dot \xi}\right),  &&
{\cal L}_{pot} = W(x)z + W_x(x)\psi\xi.
\eea
They are both manifestly supersymmetric invariant (up to a time derivative), being given by
\bea
{\cal L}_{kin} = -\frac{1}{2}Q_1Q_2 ( \psi\xi), &&
{\cal L}_{pot} = Q_1Q_2(\Phi(x))= \Phi_xz+\Phi_{xx}\psi\xi.
\eea
The second equation implies the  identification $W(x)=\Phi_x(x)$.\par
After solving the $z= - W(x)$ algebraic equation of motion for $z$,
the Lagrangian ${\cal L}$ can be expressed as
\bea\label{n2lagr}
{\cal L} &=& \frac{1}{2} ({\dot x}^2 -\psi{\dot\psi}-\xi{\dot\xi})-\frac{1}{2}W(x)^2+W_x\psi\xi.
\eea

The alternative formulation that we are presenting here can be obtained by expressing the ${\cal N}=2$ invariant action in terms
of a sigma-model Lagrangian ${\cal L}_\sigma$ plus a linear in $z$ potential term ${\cal L}_{lin}$. They are
\bea
{\cal L}_{\sigma}= Q_1Q_2(f(x)z), && {\cal L}_{lin} = \frac{1}{2}\mu z.
\eea
The total Lagrangian is
\bea\label{totallag}
{\cal L}&=& {\cal L}_\sigma+{\cal L}_{lin}=\frac{1}{2}\phi(x)({\dot x}^2+z^2-\psi{\dot\psi}-\xi{\dot\xi})+\frac{1}{2}\phi_xz\psi\xi+\frac{1}{2}\mu z,\quad {\textrm{for}} \quad \phi(x)= 2f_x.
\eea
The algebraic equation of motion for $z$ gives 
\bea
z&=& -\frac{1}{2\phi}\mu-\frac{\phi_x}{2\phi}\psi\xi.
\eea
By substituting the right hand side into the Lagrangian we obtain
\bea
{\cal L}&=& \frac{1}{2}\phi({\dot x}^2-\psi{\dot\psi}-\xi{\dot\xi})-\frac{1}{8\phi}\mu^2-\frac{\phi_x}{4\phi}\mu\psi\xi.
\eea
By performing non-linear transformations on the component fields, we can express the Lagrangian
in the so-called ``constant kinetic term" basis \textcolor{black}{\cite{HoTo}}. We set
\bea\label{barredfieldsapp}
&{\overline x}= C(x),\quad
{\overline\psi} = C_x\psi,\quad 
{\overline \xi}= C_x \xi,& \qquad {\textrm{where}}\quad
C_x= \sqrt{\phi}.
\eea
We then get, at first, the intermediate expression
\bea\label{intermediate}
{\cal L}&=& \frac{1}{2}({\dot {\overline x}}^2-{\overline\psi}{\dot{\overline\psi}}-{\overline \xi}{\dot{\overline\xi}})
-\frac{1}{8}
(\frac{\mu}{C_x})^2-\frac{\mu}{2}\frac{C_{xx}}{(C_x)^3} {\overline \psi}{\overline \xi}.
\eea
The position
\bea
&W({\overline x})=W(C(x))=\frac{\mu}{2C_x(x)}&
\eea
allows to identify (by replacing the fields $x,\psi,\xi$ with their respective barred expressions) the Lagrangian (\ref{intermediate}) with the Lagrangian (\ref{n2lagr}):
\bea
{\cal L}&=& \frac{1}{2}({\dot {\overline x}}^2-{\overline\psi}{\dot{\overline\psi}}-{\overline \xi}{\dot{\overline\xi}})-\frac{1}{2}W^2({\overline x})+W_{\overline x}{\overline\psi}{\overline \xi}.
\eea
As an example of the construction, the harmonic oscillator and the inverse-square potentials are respectively
recovered from\par
{\it i}) the harmonic oscillator potential, $W^2({\overline x})=A^2{\overline x}^2$, so that
\bea
&W({\overline x}) =A{\overline x},\qquad
C(x) =\sqrt{\frac{\mu x}{A}}, \qquad  \phi (x) = \frac{\mu}{4Ax};&
\eea

{\it ii}) the inverse square potential, $W^2({\overline x})=(\frac{g}{{\overline x}})^2$, so that
\bea
&W({\overline x}) = \frac{g}{{\overline x}},\qquad
C(x) =e^{\frac{\mu x}{2g}}, \qquad \phi = \frac{\mu^2}{4g^2}e^{\frac{\mu x}{g}}.&
\eea

%
%
%

\end{document}